\documentclass[
  aps,
  prapplied,
  twocolumn,
  superscriptaddress,
  nofootinbib
]{revtex4-2}

\usepackage{graphicx}
\usepackage{amsmath}
\usepackage{amssymb}
\usepackage{booktabs}
\usepackage{siunitx}
\usepackage{xcolor}
\usepackage{hyperref}
\usepackage{url}

\hypersetup{colorlinks=true,linkcolor=blue,citecolor=blue,urlcolor=blue}
\begin{document}

\title{Auditing Frozen-Encoder Anomaly Detection Across Mechanical Systems:\\
Representation Provenance, Calibration, and Protocol Effects}

\author{Jose S\'anchez Andreu}
\affiliation{Independent Researcher, Murcia, Spain}

\date{August 31, 2026}

\begin{abstract}
This version reports a reproducibility audit of the frozen-encoder experiments
presented in version 1. The numerical discrimination results are reproducible
from the preserved artifacts, but their original attribution to
interferometric pretraining is not supported. The released checkpoint contains
a nested model state that loads without missing parameters, whereas loading the
outer checkpoint dictionary leaves almost the entire EfficientNet-B0 feature
stack uninitialized. Preserved embeddings labelled as interferometric have
norms of order $10^{-12}$, matching freshly initialized EfficientNet-B0
networks and differing by more than twelve orders of magnitude from the
preserved ImageNet embeddings. A second, separately preserved near-zero
embedding set produces almost the same IMS 4th-test anomaly scores
($r=0.987$) and record-level discrimination (AUC $0.9812$ versus $0.9818$).

We therefore withdraw the causal claim that IMS performance demonstrates a
morphological prior transferred from gravitational-wave instrumentation. We
reanalyse the controlled IMS splits at matched observed false-positive rates
and add multivariate classical signal baselines. The near-zero representations
retain strong tail separation, particularly in the 2nd and 4th IMS runs, but
this is now interpreted as an exploratory architecture-and-initialization
effect coupled to Mahalanobis scoring. A separate PRONOSTIA audit shows that
the original large warning times were induced by a lifetime-fraction baseline;
under fixed-time evaluation, a ten-feature classical baseline outperforms the
preserved encoder scores. These results illustrate how checkpoint provenance,
finite-sample calibration, architecture, and target-domain baselines can create
an appearance of cross-domain transfer. They also define the controls required
before assigning physical meaning to frozen-representation anomaly scores.
\end{abstract}

\maketitle

\begin{widetext}
\noindent\fcolorbox{black}{gray!8}{%
\parbox{0.96\textwidth}{
\textbf{Revision statement.}
Version 1 attributed the reported anomaly separation to an EfficientNet-B0
encoder trained on interferometric transients. A subsequent internal audit
found a representation-provenance discrepancy, unequal observed false-positive
rates, and a retrospective timing artifact. The numerical values released with
version 1 remain reproducible from the preserved artifacts, but the
LIGO-specific causal interpretation and the broad physical-transfer claims are
withdrawn. This version replaces them with an artifact-level audit and reports
all known limitations. The original IMS embedding generator is unavailable;
consequently, the precise historical execution path is inferred from, rather
than proven by, the preserved artifacts and loading logs.}}
\end{widetext}

\section{Introduction}

Frozen encoders are increasingly used as generic measurement maps: an input
$x$ is transformed into a high-dimensional representation $z=F(x)$, and a
simple downstream statistic is used to detect novelty or change. This strategy
is attractive when target-domain labels are scarce, but it creates a demanding
causal question. Strong performance of a frozen pipeline does not by itself
show that source-domain training supplied the useful prior. The effect may
instead arise from architecture, random projection, preprocessing, target
calibration, or the downstream detector.

Version 1 of this work proposed that a representation trained on
gravitational-wave detector transients transferred a morphology-sensitive
prior to mechanical vibration signals. The strongest evidence came from an
IMS--NASA controlled split, where the representation labelled ``LIGO'' reached
record AUC $0.982$ and a reported tail-enrichment factor of $859.5$ on the 4th
run, while an ImageNet-pretrained EfficientNet-B0 reached AUC $0.652$ and tail
enrichment $1.97$. Run-to-failure experiments were also interpreted as early
warnings of progressive degradation.

The present version asks a narrower question: what do the preserved artifacts
actually establish? We audit (i) checkpoint loading, (ii) embedding scale and
geometry, (iii) realized rather than nominal false-positive rates, (iv)
classical signal baselines, and (v) the chronology of the prognostic protocol.
Our objective is not to recover the original interpretation, but to distinguish
reproducible numerical effects from unsupported causal attribution.

The audit yields three principal conclusions. First, the preserved embeddings
labelled as interferometric are overwhelmingly consistent with untrained random
features, although the missing historical generator prevents absolute proof of
the exact execution route. Second, the resulting anomaly separation on selected
IMS splits is real at the artifact level and survives observed-FPR matching,
but it cannot be attributed to interferometric training. Third, the original
run-to-failure warning times were not prospective estimates of incipient fault;
they mostly reflected the construction of the early-life baseline.

\section{Audit materials and methods}

\subsection{Preserved artifacts and numerical reproduction}

The audit uses the window-level Parquet artifacts and metric summaries released
for the IMS controlled splits, the corresponding public IMS raw signals, the
preserved PRONOSTIA embeddings and score tables, and the archived Aether
checkpoint. No abnormal target-domain sample is used to fit the anomaly model.

For each of three IMS runs, the controlled split contains 120 early-life
reference records, 250 held-out nominal records, and 250 late-life records.
Each record contributes 31 overlapping windows. We reproduced the version-1
window AUC, record AUC, threshold exceedance counts, and tail-enrichment values
directly from the artifacts. Equality with the released values was used as a
guard against score-column, label, split, or aggregation errors.

\subsection{Checkpoint-loading audit}

The archived checkpoint has four top-level entries:
\texttt{model\_state\_dict}, \texttt{class\_names}, \texttt{config}, and
\texttt{metrics}. We tested two plausible loading routes with the current
\texttt{torchvision} EfficientNet-B0 implementation. Passing the outer
dictionary directly to \texttt{load\_state\_dict(..., strict=False)} yields 311
missing model keys and four unexpected top-level keys. Loading the nested
\texttt{checkpoint["model\_state\_dict"]} yields zero missing and zero
unexpected keys. An archived execution note reports the closely corresponding
historical count of 309 missing and four unexpected keys, consistent with a
minor library-version difference.

The exact IMS embedding-generation script is absent. We therefore distinguish
two statements throughout: the checkpoint structure and load counts are direct
observations, whereas attribution of a particular preserved embedding file to
the incorrect route is a forensic inference.

\subsection{Embedding fingerprint}

For each preserved embedding $z\in\mathbb{R}^{1280}$ we compute its Euclidean
norm and coordinate scale. We compare these values with outputs from freshly
initialized EfficientNet-B0 models and from the correctly loaded checkpoint on
controlled synthetic inputs. We additionally compare two preserved IMS 4th-run
near-zero embedding sets, one labelled ``LIGO'' and one without an encoder
label, using score correlations, centered linear kernel alignment (CKA), and
pairwise cosine-geometry correlation.

\subsection{Anomaly scores and operating points}

The preserved anomaly score is the distance-form Mahalanobis statistic
\begin{equation}
S(z)=\sqrt{(z-\mu_0)^\top\Sigma_{\mathrm{LW}}^{-1}(z-\mu_0)},
\end{equation}
where $\mu_0$ and the Ledoit--Wolf covariance estimate
$\Sigma_{\mathrm{LW}}$ are fitted on the early-life reference set
\cite{ledoit2004}. Record-level AUC uses the within-record 0.95 score quantile.

Version 1 defined $\tau=Q_{0.999}(S\mid\mathcal P_0)$ on the 120 reference
records and described this convention as a common $0.1\%$ false-positive rate.
This is a calibration target, not a guarantee on held-out nominal records. We
therefore report both (i) transfer of the original $\mathcal P_0$ threshold and
(ii) a post-hoc diagnostic in which each method is recalibrated to the same
observed $0.1\%$ held-out nominal rate. The latter is not presented as a
prospective detector; it isolates ranking and tail separation at a common
operating point.

Uncertainty is estimated by a paired 5000-sample cluster bootstrap over whole
records. The nominal and abnormal record indices are resampled separately, and
the observed-FPR threshold is re-estimated within every bootstrap sample.

\subsection{Classical IMS baselines}

Raw public IMS channel-0 signals are resampled to 4096 Hz and divided into the
same 256-sample windows with 128-sample overlap. A ten-dimensional baseline
contains RMS, peak magnitude, crest factor, kurtosis, skewness, standard
deviation, mean absolute value, envelope RMS, normalized spectral entropy, and
spectral bandwidth. Features are robustly scaled from the reference windows,
and the same Ledoit--Wolf Mahalanobis detector is fitted. We also evaluate RMS
and kurtosis separately. Two resampling variants, using either all 20480 raw
samples or the first 20000, give materially identical results.

\subsection{PRONOSTIA timing audit}

The historical PRONOSTIA protocol defined warm-up as the first $5\%$ of each
bearing lifetime and the baseline as the following $15\%$. Thus, the end of the
reference interval depended on the known end-of-life time. We recover warning
and alarm times for all 17 bearings and report their positions relative to the
baseline and total lifetime.

For a non-fractional diagnostic, reference intervals are fixed in elapsed time.
The preserved Aether analysis uses the same nine validation bearings for which
embeddings are available; it is compared with the ten-feature classical
Mahalanobis baseline. A detection is counted as operationally successful only
if no false-alarm episode occurs in the early validation interval and a later
persistent detection occurs. PRONOSTIA provides no independent physical
fault-onset annotation, so the fixed intervals remain operational proxies, not
certified healthy and faulty states.

\subsection{Preprocessing dimensionality check}

The documented pipeline segments a signal into 256-sample windows and applies a
256-point Hann STFT with \texttt{center=False} to each such window. This produces
129 frequency bins and one temporal frame. Resizing this $129\times1$ array to
$224\times224$ does not create within-window temporal information. Accordingly,
results generated by this documented pipeline can be interpreted as sensitivity
to normalized spectral shape, but not as direct evidence of within-window
time--frequency organization or phase sensitivity.

\section{Results}

\subsection{The version-1 numerical artifacts are internally reproducible}

The headline IMS values are reproduced exactly from the preserved score files.
For the representation labelled ``LIGO'', the record AUC values are $0.7016$,
$1.0000$, and $0.9818$ for the 1st, 2nd, and 4th runs, respectively. The
corresponding version-1 tail-enrichment values are $1.27$, $191.6$, and $859.5$.
Thus, the audit does not allege numerical fabrication. It changes the
interpretation of those numbers and the validity of their operating-point
comparison.

\subsection{The preserved ``LIGO'' embeddings have an untrained-network fingerprint}

Table~\ref{tab:fingerprint} summarizes the preserved IMS 4th-run embedding
scales. Every coordinate in both near-zero sets has absolute magnitude below
$10^{-10}$. Their median norms are $1.64\times10^{-12}$ and
$4.23\times10^{-12}$, while the ImageNet embeddings have median norm $15.22$.
Freshly initialized EfficientNet-B0 models evaluated in inference mode also
produce pooled activations of order $10^{-12}$ on nonzero controlled inputs.
The correctly loaded trained checkpoint does not share this fingerprint: even
zero and constant inputs give norms near 7, while generic random images produce
much larger responses.

\begin{table}[t]
\caption{Preserved IMS 4th-run embedding fingerprint. ``Near-zero A'' is the
unlabelled default artifact and ``Near-zero B'' is the artifact labelled LIGO.}
\label{tab:fingerprint}
\resizebox{\linewidth}{!}{%
\begin{tabular}{lccc}
\toprule
Representation & Median $\lVert z\rVert_2$ & 5--95\% range & $|z_j|<10^{-10}$ \\
\midrule
Near-zero A & $1.64\times10^{-12}$ & $(1.13$--$2.03)\times10^{-12}$ & 100\% \\
Near-zero B & $4.23\times10^{-12}$ & $(2.93$--$5.30)\times10^{-12}$ & 100\% \\
ImageNet & $15.22$ & $14.17$--$16.37$ & 0\% \\
\bottomrule
\end{tabular}}
\end{table}

\begin{figure*}[t]
\centering
\includegraphics[width=0.96\textwidth]{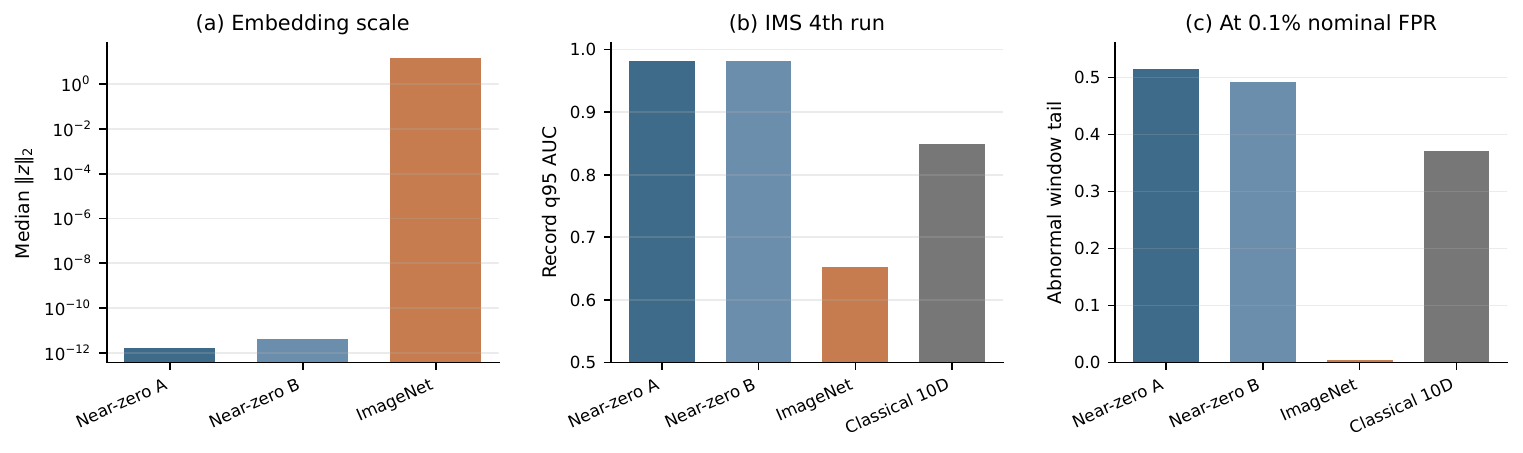}
\caption{\textbf{Representation fingerprint and IMS 4th-run performance.}
(a) Median embedding norm on a logarithmic scale. (b) Record-level AUC from the
within-record 0.95 score quantile. (c) Abnormal window tail occupancy after
post-hoc calibration to a common observed nominal false-positive rate of
$0.1\%$. The classical baseline uses ten raw-signal descriptors. Near-zero A
and B are separately preserved artifacts, both consistent with untrained
EfficientNet-B0 activation scale. Error bars are omitted here because the
figure compares point estimates; cluster-bootstrap intervals are reported in
the text and audit tables.}
\label{fig:fingerprint}
\end{figure*}

The two near-zero sets are not numerically identical, but their anomaly scores
are strongly aligned: Pearson $r=0.9875$ and Spearman $\rho=0.9829$ across
19220 windows. On fixed random subsets, centered linear CKA is $0.806$ and the
pairwise cosine-geometry correlation is $0.840$. Most importantly, they produce
almost the same IMS 4th-run outcome: record AUC $0.9812$ and $0.9818$, and
abnormal tail occupancy $0.5147$ and $0.4921$ at the same observed $0.1\%$
nominal rate. Figure~\ref{fig:fingerprint} summarizes these contrasts.

Taken together with the checkpoint structure and archived missing-key log,
these observations provide strong evidence that the preserved result did not
measure the intended trained encoder. Because the original generator is absent,
we stop short of claiming proof of the precise software path.

\subsection{The nominal calibration rate did not transfer equally}

The $Q_{0.999}(\mathcal P_0)$ threshold produces four exceedances among 3720
reference windows for every encoder/run combination. Held-out nominal
exceedance rates, however, range from $0.0516\%$ to $16.58\%$ for the near-zero
representation and from $2.05\%$ to $4.57\%$ for ImageNet. Table
\ref{tab:historical} shows that the large version-1 enrichment ratio is partly
controlled by the held-out denominator. In the 4th run, $859.5$ is exactly the
ratio of 3438 abnormal exceedances to four held-out nominal exceedances after
equal window counts are taken into account.

\begin{table*}[t]
\caption{Reproduction of the original controlled-split comparison. FPR is the
held-out nominal exceedance rate under the threshold estimated from the
reference set. Classical AUC uses the ten-feature baseline; it was not part of
the version-1 enrichment calculation.}
\label{tab:historical}
\begin{ruledtabular}
\begin{tabular}{lcccccc}
IMS run & Near-zero AUC & ImageNet AUC & Classical AUC & Near-zero held-out FPR & ImageNet held-out FPR & Version-1 $\Lambda$ \\
\hline
1st & 0.7016 & 0.5378 & 0.7160 & 16.58\% & 4.57\% & 1.27 \\
2nd & 1.0000 & 0.9974 & 0.9959 & 0.503\% & 2.05\% & 191.6 \\
4th & 0.9818 & 0.6520 & 0.8487 & 0.0516\% & 3.52\% & 859.5 \\
\end{tabular}
\end{ruledtabular}
\end{table*}

Consequently, the statement that both encoders were compared under the same
fixed false-alarm rate is incorrect. They used the same reference-quantile
convention but experienced different false-positive rates after transfer to
held-out nominal records.

\subsection{Observed-FPR matching preserves an artifact-level effect}

Table~\ref{tab:equalfpr} reports the post-hoc common-operating-point diagnostic.
At an observed nominal rate of $0.0903\%$ (seven exceedances among 7750 nominal
windows), abnormal tail occupancy remains much larger for the near-zero
representation than for ImageNet in all three runs. The paired cluster
bootstrap gives near-zero mean tail occupancies and 95\% intervals of
$0.0975$ [0.0641, 0.1326], $0.7788$ [0.7064, 0.8699], and $0.4913$
[0.4227, 0.5521] for the 1st, 2nd, and 4th runs.

\begin{table}[t]
\caption{Abnormal window-tail occupancy at a common observed nominal FPR of
approximately $0.1\%$. Values are point estimates.}
\label{tab:equalfpr}
\begin{ruledtabular}
\begin{tabular}{lccc}
IMS run & Near-zero & ImageNet & Classical \\
\hline
1st & 0.0982 & 0.00168 & 0.00452 \\
2nd & 0.7693 & 0.0693 & 0.3768 \\
4th & 0.4921 & 0.00452 & 0.3712 \\
\end{tabular}
\end{ruledtabular}
\end{table}

Compared with the classical baseline, the bootstrapped mean tail differences
for near-zero minus classical are $0.0881$ [0.0543, 0.1250], $0.3701$
[0.1014, 0.5653], and $0.1212$ [0.0471, 0.1823]. These are genuine
artifact-level separations. They do not identify interferometric training as
the cause. They motivate a new, explicitly exploratory hypothesis: random
convolutional features combined with covariance-normalized distance can amplify
particular changes in normalized one-frame spectral shape.

At the record level, the near-zero and classical baselines are statistically
tied in the 1st run: AUC difference $-0.0144$, 95\% bootstrap interval
[-0.0662, 0.0387]. The near-zero representation is slightly higher in the 2nd
run (difference $0.00414$ [0.00035, 0.00943]) and clearly higher in the 4th
run (difference $0.1331$ [0.1004, 0.1703]). Single-feature baselines show that
the runs are physically heterogeneous: RMS alone reaches record AUC $1.000$ in
the 2nd run and $0.913$ in the 4th, whereas kurtosis reaches $0.681$ in the 1st
but only $0.393$ in the 4th.

\subsection{PRONOSTIA warning time was induced by the baseline design}

In the historical analysis, the baseline ended at $20\%$ of each bearing's
known lifetime. The median warning and alarm positions were $20.13\%$ and
$20.19\%$ of life. Fifteen of 17 warnings occurred within one minute after the
baseline ended, and all 17 occurred within five minutes. The reported median
lead was 190 minutes, but total lifetime and reported lead had Spearman
$\rho=1.000$. This near-deterministic relation shows that the large lead time
primarily measured the remaining lifetime after an alarm placed close to the
retrospectively defined $20\%$ boundary.

\begin{figure}[t]
\centering
\includegraphics[width=\columnwidth]{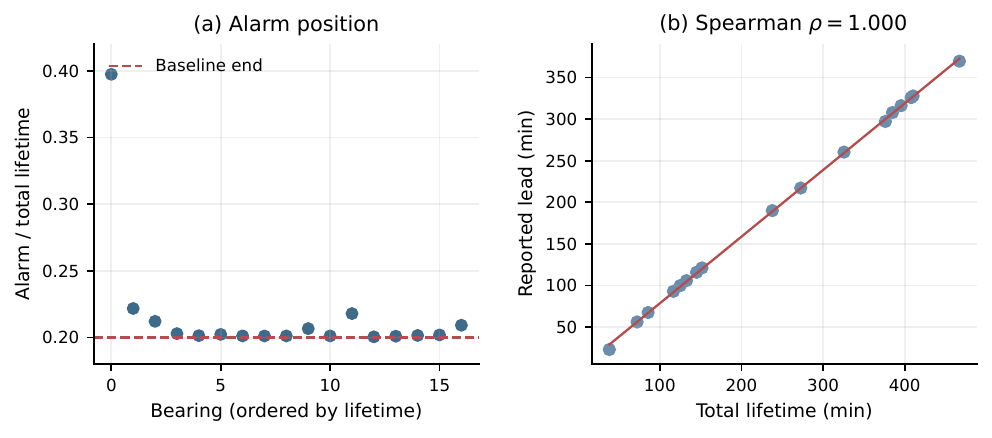}
\caption{\textbf{PRONOSTIA historical timing diagnostic.}
(a) Alarm position as a fraction of total lifetime. The dashed line marks the
end of the retrospectively defined baseline at $20\%$. (b) Reported lead time
versus total lifetime. Their rank correlation is exactly one, showing that the
lead statistic is dominated by the lifetime-dependent baseline construction.}
\label{fig:pronostia}
\end{figure}

Under the fixed-time target-calibrated diagnostic on the nine validation
bearings with preserved embeddings, the Aether score has no early false-alarm
episodes, six operational successes, median successful lead 1.5 minutes, and
median temporal Spearman correlation $0.471$. The matched classical baseline
has no early false-alarm episodes, nine successes, median lead 74 minutes, and
median correlation $0.969$. Under a stricter external-series Aether reference,
the mean early false-positive rate rises to $69.0\%$, eight of nine bearings
show false-alarm episodes, and only one bearing meets the operational-success
criterion. Table~\ref{tab:pronostia} summarizes the aligned primary comparison.

\begin{table}[t]
\caption{Fixed-time PRONOSTIA diagnostic on the same nine validation bearings.
Lead is reported only among operational successes.}
\label{tab:pronostia}
\resizebox{\linewidth}{!}{%
\begin{tabular}{lcccc}
\toprule
Method & Success & False alarms & Median lead & Median $\rho$ \\
\midrule
Preserved Aether & 6/9 & 0/9 & 1.5 min & 0.471 \\
Classical 10D & 9/9 & 0/9 & 74.0 min & 0.969 \\
\bottomrule
\end{tabular}}
\end{table}

These results do not show that the preserved Aether score is devoid of trend:
all nine correlations are positive and pass the circular-shift false-discovery
control. They show instead that the trend is neither encoder-specific nor a
competitive prospective warning signal under the tested protocol.

\section{Discussion}

\subsection{Claims withdrawn from version 1}

The audit does not support the following version-1 claims: (i) that the IMS
separation demonstrates a prior learned from interferometric transients; (ii)
that the ImageNet comparison isolates source-domain training as the causal
factor; (iii) that the reported tail ratios compare encoders at the same
observed false-positive rate; (iv) that the historical PRONOSTIA lead times
constitute prospective early warnings; or (v) that the documented
$129\times1$ representation directly probes within-window temporal
organization. Broader statements about universal or physically selective
cross-system morphological transfer are therefore withdrawn.

\subsection{What remains empirically interesting}

The correction does not make every numerical pattern disappear. Two preserved
near-zero feature sets produce nearly identical IMS 4th-run scores and both
substantially outperform ImageNet and the tested classical baseline in the
extreme tail. Because Mahalanobis distance is invariant to a uniform global
rescaling when covariance is estimated consistently, the small absolute
activation scale is not itself the source of performance. Rather, the useful
quantity is the relative geometry induced across windows.

Random convolutional networks can generate structured, input-dependent
representations even without supervised training. The present evidence is
compatible with that general principle, but it is insufficient to establish a
stable random-feature method: only two relevant preserved near-zero sets are
available, their exact initialization seeds are unknown, and the split was not
designed prospectively for a random-feature hypothesis. A valid follow-up must
pre-register multiple random seeds, include direct spectral kernels and random
linear projections, evaluate multiple architectures, and test independent
datasets and unscreened trajectories.

\subsection{Why the original numerical result looked persuasive}

Several individually reasonable choices combined to produce a misleading
causal narrative. A high-dimensional architecture generated a rich geometry;
a covariance-normalized detector converted that geometry into strong anomaly
scores; a reference quantile was interpreted as a held-out operating rate; and
the ImageNet baseline behaved differently. Without a random-initialization
control and verified parameter loading, the difference between two encoders
was attributed to the intended source-domain training. The audit shows that
architecture and initialization cannot be treated as passive implementation
details.

The prognostic analysis illustrates a parallel issue. Defining the reference
interval as a fraction of total lifetime appears normalized and comparable, but
it introduces end-of-life knowledge into the alarm clock. A detector that
crosses immediately after that boundary automatically inherits a lead time
proportional to lifetime. Trend and warning utility must therefore be evaluated
separately.

\subsection{Limitations of the audit}

The original IMS generator and exact software environment are unavailable. We
cannot prove the precise historical load call, reconstruct every preprocessing
detail bit-for-bit, or exclude an undocumented postprocessing transformation of
the embeddings. The intended trained checkpoint has not yet been rerun over the
full controlled IMS split with a verified generator. Accordingly, this paper
does not estimate the true IMS performance of a correctly deployed Aether
encoder.

The equal-observed-FPR analysis is deliberately post hoc and should not be
reported as prospective calibration. The controlled IMS split uses selected
early and late blocks from only three trajectories. Window observations overlap
and are not independent; record-cluster bootstrap reduces but does not remove
trajectory-level uncertainty. The classical baseline is a strong matched
diagnostic, not an exhaustive condition-monitoring benchmark. Finally,
PRONOSTIA lacks an independent physical annotation of fault onset, so neither
large nor small lead values can be interpreted as verified incipient-fault
warning times.

\section{Conclusion}

The central physical-transfer interpretation of version 1 does not survive
artifact-level reproducibility audit. The reported IMS discrimination values
are numerically real within the preserved pipeline, but the embeddings
attributed to interferometric pretraining have the fingerprint of untrained
EfficientNet-B0 features. Unequal held-out calibration and a retrospective
lifetime-fraction baseline further inflated the apparent strength of the
evidence.

The corrected result is narrower but useful. Random or near-random
convolutional geometry combined with Mahalanobis scoring can produce strong
separation on selected IMS late-life blocks, and this effect can exceed both an
ImageNet representation and a ten-feature classical baseline in the extreme
tail. Whether that observation generalizes is an open question requiring a new
pre-registered study. The broader methodological lesson is immediate:
source-domain transfer should not be inferred without verified checkpoint
provenance, random-initialization controls, matched observed operating points,
classical baselines, and chronology-preserving validation.

\section*{Data and code availability}

The raw IMS and PRONOSTIA datasets are publicly available from their original
providers. An audit supplement accompanying this version contains deterministic
scripts, preserved derived tables, checksums, per-record results, and bootstrap
summaries that permit verification of the reported artifact-level calculations.
The private source-domain training data and model weights are not redistributed;
therefore, independently rerunning checkpoint-dependent diagnostics requires
access to those weights. A cryptographic hash of the audited checkpoint and the
exact key-loading diagnostics are included so that its identity and the preserved
audit outcome are unambiguous.

\end{document}